\documentclass[AMA,STIX1COL]{WileyNJD-v2}

\articletype{Article Type}%

\received{26 April 2016}
\revised{6 June 2016}
\accepted{6 June 2016}

\begin{document}

\title{Publication bias adjustment in network meta-analysis: an inverse probability weighting approach using clinical trial registries}

\author[1]{Ao Huang}

\author[2]{Yi Zhou}

\author[3]{Satoshi Hattori}

\authormark{ Huang \textsc{et al}}

\address[1]{\orgdiv{Department of Medical Statistics}, \orgname{University Medical Center G\"{o}ttingen}, \orgaddress{\state{Humboldtallee 32, 37073 G\"{o}ttingen}, \country{Germany}}}

\address[2]{\orgdiv{Beijing International Center for Mathematical Research},\orgname{Peking University}, \orgaddress{\state{No.5 Yiheyuan Road, Haidian District, Beijing 100871}, \country{China}
}}
\address[3]{\orgdiv{Department of Biomedical Statistics, Graduate School of Medicine,\\  Integrated Frontier Research for Medical Science Division, \\
        Institute for Open and Transdisciplinary Research Initiatives (OTRI)}, \orgname{Osaka University}, \orgaddress{\state{Yamadaoka 2-2, Suita City, Osaka 565-0871}, \country{Japan}}}

\corres{Satoshi Hattori, Department of Biomedical Statistics, Graduate School of Medicine, Osaka University, Japan. \email{hattoris@biostat.med.osaka-u.ac.jp}}

\abstract[Summary]{Network meta-analysis (NMA) is a useful tool to compare multiple interventions simultaneously in a single meta-analysis, it can be very helpful for medical decision making when the study aims to find the best therapy among several active candidates. However, the validity of its results is threatened by the publication bias issue. Existing methods to handle the publication bias issue in the standard pairwise meta-analysis are hard to extend to this area with the complicated data structure and the underlying assumptions for pooling the data. In this paper, we aimed to provide a flexible inverse probability weighting (IPW) framework along with several $t$-type selection functions to deal with the publication bias problem in the NMA context. To solve these proposed selection functions, we recommend making use of the additional information from the unpublished studies from multiple clinical trial registries. A comprehensive numerical study and a real example showed that our methodology can help obtain more accurate estimates and higher coverage probabilities, and improve other properties of an NMA (e.g., ranking the interventions). 
}


\maketitle

\section{Introduction}\label{Int}
Network meta-analysis (NMA), which is also known as the mixed treatment comparisons or the multiple treatment meta-analysis, is an extension of the standard pairwise meta-analysis\cite{rouse2017,salanti2011}. In the standard pairwise MAs, we are restricted to addressing only one question in a head-to-head comparison; while in NMAs, we are supposed to answer a more complicated question that which treatment should be recommended among a set of candidate treatments. Such consideration often arises in the practice of medical decision making\cite{efthimiou2016,antoniou2023}. If we have many treatment options, it is a typical situation that there is no direct evidence for some pairs of treatments. Here, the direct evidence means the estimates obtained through the estimates from studies that had been carried out to compare two treatments directly; the indirect evidence means the estimates obtained through the studies with common comparators (e.g., the relative effect of two active treatments could be obtained indirectly through a placebo group)\cite{rouse2017}. The NMA addresses this issue by synthesizing direct and indirect evidence with a multivariate model. The main results of an NMA comprise a league table to show the comparative effectiveness among any pair of the included treatments. This can be especially useful when two competing interventions have not been directly compared in a head-to-head study\cite{balduzzi2023}. Another important output is the ranking metrics which can help us to select the best treatment for medical decision making. However, as concerned in the standard MAs, the validity of the NMA results is threatened by the publication bias issue. It is the bias caused by the tendency that the studies with statistically significant results are more likely to be published in scientific journals. Due to the complexity of the data structure and underlying assumptions behind the model, (e.g., the consistency assumption which means all the evidence obtained directly or indirectly should be no big difference, the transitivity assumption which means common comparators should be similar across different trials), the publication bias issue becomes much more complicated for the NMA. Although there have been many methods against publication bias in the pairwise MA context, such as the funnel plot, regression-based tests, and sensitivity analysis methods with a selection function, a limited number of methods have been developed for the NMA. 

The funnel plot is a widely-used graphical approach in the pairwise MAs \cite{jin2015}. For the NMA, Chaimani et al. \cite{cha2013} proposed the comparison-adjusted funnel plot. It is an extension of the standard funnel plot that uses the centered treatment effects (study-specific effect sizes minus the comparison-specific summary effect) against the measure of precision. Then the selective publication could be addressed with funnel plot asymmetry around the zero line, and those graph-based tests (e.g., Egger's regression test\cite{egger1997}, Begg's rank correlation test\cite{begg1994}) could be employed as well. However, such kind of graphical methods are known to be subjective and have low power and type I error inflation, since the selective publication process is not the only reason for the funnel plot asymmetry\cite{jin2015,marks2020}.

Modeling the selective publication process with a selection function is a more objective alternative to address the publication bias issue\cite{jin2015,marks2020}. It can quantify the impacts of the publication bias and provide the corrected results. The selection function describes the probability that a study with a specific outcome is published. The Copas-Shi selection function is one of the most widely used methods for this purpose\cite{copas2000,copas2001}, which was originally proposed in the area of econometrics by Heckman \cite{heckman1979} to solve the sample selection problem. The Copas-Shi selection function is defined as a latent Gaussian model. It successfully introduced the selective publication of dependence with the outcome as well as the study size. However, it is not necessarily easy to interpret and understand the selective publication process under consideration. Alternatively, there are several $t$-type selection functions that model the selective publication process with $t$-statistics used in each study\cite{preston2004,copas2013}. Since the $P$-values were deemed as an influential factor in the selective publication process\cite{moreno2009}, the idea of modeling the selective publication process as a function of the $t$-statistics might be more appealing and interpretable. Recently, a few papers proposed sensitivity analysis methods for publication bias in the NMAs. All of them relied on the Heckman-type selection functions \cite{marks2022,mavridis2013,mavridis2014selection}. In this paper, we would like to address the publication bias issue with several $t$-type selection functions. We take the approach proposed by Huang et al. utilizing the clinical trial registries to solve the parameter estimation problems\cite{huang2023}.

The rest of the paper is organized as follows. In Section 2, we start by introducing the general data structure in an NMA and then illustrate the standard approach of the multivariate random-effects model. In Section 3, we introduce the expanded data structure and our proposed IPW approach with clinical trial registries. We present simulation results to show the performance of our proposed method in Section 4. Subsequently, we applied our proposed method to a real NMA example in Section 5, which was known to suffer from the publication bias issue. We conclude our paper by addressing some remaining issues and future works in Section 6.

\section{The standard framework for network meta-analysis}
\subsection{Multivariate random-effects model for network meta-analysis }
Suppose we are interested in comparing treatment effects of $T$ treatments (${A,B,C\ldots}$). We conduct a comprehensive literature search to identify all the relevant studies using these treatments and then synthesize their reported outcomes with an NMA. Differently from the standard pairwise MAs, an NMA may involve multi-arm trials as well as pairwise trials. To distinguish these studies, following Mavridis et al.\cite{mavridis2014selection}, we introduce the index $k$, which represents the type of the treatment groups involved, and the set of the treatments is denoted by $T_k\subseteq\left\{A,B,C\ldots\right\}$, where $k=1,2,\ldots,K$. Suppose each study design consists of $N_k$ studies, then the total number of studies included in this NMA equals $N=\sum_{k=1}^K N_k$.

For simplicity, we describe our proposal assuming that there are three treatments ($T=3$) and the number of the design type is $K=4$. It involves two active treatments $A$ and $B$ and a control group of placebo $C$. Extension to the case of more treatment groups can be done straightforwardly. The corresponding study design $k$ in this dataset consists of 4 types: $T_1=\left\{A,C\right\}$,\ $T_2=\left\{B,C\right\}$,\ \ $T_3=\left\{A,B\right\}$,\ $T_4=\left\{A,B,C\right\}$. Let $y_{ik}^{XY}$ denote the estimated treatment effect (e.g., log-odds ratio, log-hazard ratio) of $i$th study in design $k$ for comparing treatments $X$ and $Y$. For example, $y_{i4}^{AC}$ means the observed relative treatment effect of $A$ versus $C$ from $i$th study in the multi-arm trial design 4. Similarly, let $s_{ik}^{XY}$ and $n_{ik}^{XY}$ be the standard error of $y_{ik}^{XY}$ and the number of subjects for the comparison for the treatments $X$ and $Y$ in the $i$th study of the design $k$. In a multi-arm trial, we need the covariance of the treatment effects within a study. Suppose our interest is to estimate the relative treatment effects of active treatments $A$ and $B$ versus the placebo $C$, which were denoted by $\mu^{AC}$ and $\mu^{BC}$, respectively (the relative effect of $A$ versus $B$ can be estimated as $\mu^{AB}=\mu^{AC}-\mu^{BC}$ according to the consistency assumption). To synthesize the aforementioned data, we introduce the standard multivariate random-effects model (MRE) as follows. For the observed outcomes from the two-arm trial designs ($k$=1, 2 and 3), similar to the pairwise MAs,
we assume the random-effects models as
\begin{eqnarray}
y_{i1}^{AC}\sim N(\mu_{i1}^{AC},(s_{i1}^{AC})^2),\
\mu_{i1}^{AC}\sim N(\mu^{AC},\tau_1^2)  \\
y_{i2}^{BC}\sim N(\mu_{i2}^{BC},(s_{i2}^{BC})^2),\
\mu_{i2}^{BC}\sim N(\mu^{BC},\tau_2^2)\nonumber \\
y_{i3}^{AB}\sim N(\mu_{i3}^{AB},(s_{i3}^{AB})^2),\
\mu_{i3}^{AB}\sim N(\mu^{AB},\tau_3^2),\nonumber
\end{eqnarray}
where $\mu_{ik}^{XY}$ is a random-effect following a normal distribution with mean $\mu^{XY}$ and design-specific between-study heterogeneity $\tau_k$.

For the observed outcomes from the three-arm design ($k$=4), we assume the bivariate random-effects model as
\begin{equation}
\begin{gathered}
\begin{pmatrix} 
y_{i4}^{AC} \\ y_{i4}^{BC}
\end{pmatrix}
\end{gathered}
\sim
\begin{gathered}
N\begin{pmatrix}
\begin{gathered}
\begin{pmatrix} \mu_{i4}^{AC} \\ \mu_{i4}^{BC} \end{pmatrix} , \begin{pmatrix} (s_{i4}^{AC})^2 & cov(y_{i4}^{AC},y_{i4}^{BC})\\  cov(y_{i4}^{AC},y_{i4}^{BC})& (s_{i4}^{BC})^2\end{pmatrix}
\end{gathered}
\end{pmatrix}
\end{gathered},
\end{equation}
where
\begin{equation}
\begin{gathered}
\begin{pmatrix} \mu_{i4}^{AC} \\ \mu_{i4}^{BC} \end{pmatrix}
\end{gathered}\sim \begin{gathered}
N\begin{pmatrix}
\begin{gathered}
\begin{pmatrix}\mu^{AC} \\ \mu^{BC} \end{pmatrix},\begin{pmatrix}\tau_4^2 & \tau_4^2/2 \\ \tau_4^2/2 & \tau_4^2 \end{pmatrix}
\end{gathered}
\end{pmatrix}
\end{gathered}
\end{equation}
and $cov(y_{i4}^{AC},y_{i4}^{BC})$ is the covariance between the comparison of $AC$ and $BC$. It agrees with the variance of the shared group $C$ and then is assumed to be extracted from the paper. To have a simple expression of the log-likelihood function, we introduce the following notation. For two-arm trials ($k=1,2$ and $3$), set $\mathbf{\Sigma_{ik}}=s_{i1}^{AC}$, $s_{i2}^{BC}$, and $s_{i3}^{AB}$ in (1), depending on the corresponding groups to compare. That is, $\mathbf{\Sigma_{ik}}$ is a scalar. For multi-arm trials (k=4), $\mathbf{\Sigma_{ik}}$ is defined as the variance-covariance matrix in the model (2). In this case, it is a matrix. With these notations,
the log-likelihood of the whole network can be written as
\begin{align}
    logL(\mathbf{\mu,\tau})=-\dfrac{1}{2}\sum_k^4\sum_{i=1}^{N_k} \left\{log\mid \mathbf{\Omega_{ik}+\Sigma_{ik}}\mid+(\mathbf{y_{ik}-\mu_k})^T(\mathbf{\Omega_{ik}+\Sigma_{ik}})^{-1}(\mathbf{y_{ik}-\mu_k})\right\},
    \label{mle}
\end{align}
where $\mathbf{\mu_k}$ refers to the overall treatment effect(s) within each study design, $\mathbf{\Omega_{ik}}$ denote the corresponding heterogeneity structure of $i$th study in design $k$, it was a scalar when $k$=1, 2 and 3, and a matrix when $k$=4 as shown in Model (3). To estimate parameters of interest ($\mathbf{\mu_k}$, $\mathbf{\Omega_{ik}}$), we can use the standard maximum likelihood method by maximizing the log-likelihood function ($\ref{mle}$). The variance of these parameters can be calculated with the resulting hessian matrix, and one can construct the 95\% confidence interval following the standard asymptotic theory for the maximum likelihood estimation.

\subsection{Ranking treatments with the $P$-score method}
An important feature of the NMA is it gives us an easy way to understand the superiority across different interventions by ranking the treatment effects.
There are several methods available for ranking the treatments from different perspectives. Salanti et al.\cite{salanti2011} developed the summary measure of the Surface Under
the Cumulative RAnking (SUCRA) in a Bayesian framework. R\"{u}cker et al.\cite{rucker2015} introduced the so-called $P$-score method which is a frequentist analogue to SUCRA. Both of these two methods are based on the probability that one treatment is better than the rest treatments. Other ranking methods have been developed based on very similar ideas to the $P$-score method\cite{nikolakopoulou2021,chaimani2021,brignardello2018,chiocchia2020}, we focus on the original $P$-score method in this paper. Let
 \begin{eqnarray}
        P_{ij} & = & \left\{\begin{array}{ll} \Phi\Biggl(\dfrac{\hat{\mu}_{ij}}{\hat{\sigma}_{ij}}\Biggl)& \mbox{if the benefit is supposed to be positive} \\ 1-\Phi\Biggl(\dfrac{\hat{\mu}_{ij}}{\hat{\sigma}_{ij}}\Biggl) & \mbox{if the benefit is supposed to be negative}\end{array}\right.\mbox{,} \label{pscore}
      \end{eqnarray}
      
where$\hat{\mu}_{ij}$ is the estimated difference of the outcomes between the $i$th and $j$th treatments, and $\hat{\sigma}_{ij}$ is its standard error. Then $P_{ij}$ is the one-sided $p$-value for the $t$-statistic of them. The $P$-score for the treatment $i$ is defined as $\bar{P_i}=\dfrac{1}{T-1}\sum_{i,i\neq j}^{T}P_{ij}$, which simply means the mean extent of certainty that the treat $i$ is better than any other treatment $j$. The treatments are then ranked according to $P$-score.

\section{Proposed methods}
\subsection{Data extension with Clinical trial registries}
Prospective registration of study protocols to clinical trial registries could be regarded as a useful non-statistical approach for the publication bias issue. It allows systematic reviewers to identify all the eligible studies including those unpublished studies and their corresponding information (e.g., planned sample size), sometimes even their outcomes are also available. Since the International Committee of Medical Journal Editors (ICMJE) initiated the trial-registration policy as a requirement for publication in its member journals in July 2005 \cite{deangelis2005}, more and more countries and regions started to promote the legislation on trial registration\cite{bian2010}. Nowadays several clinical trial registries have been well-established (for details of the clinical trial registries, see Viergever et al.\cite{viergever2015}) and widely accepted by the trialists. Research showed that the accumulated information from these registries could be a very useful resource to reduce publication bias \citep{hart2012,baudard2017,alqaidoom2022}. However, these registries mostly served as a search tool in MA practice. Some recent papers proposed methods to quantify publication bias efficiently utilizing information in clinical trial registries. Although registered information is not consistent in clinical trial registries, the planned sample size is commonly registered. The standard dataset for pairwise MA can be expanded by attaching the planned sample size of all the published and unpublished studies, and the extended dataset looks like the dataset arising from the standard missing data problem. Then, the methods for the missing data analysis can be translated into the MA. Instead of maximizing the conditional likelihood function given published, Huang et al.\cite{huang2021} handled the sensitivity analysis method with the Copas-Shi selection function more stably by using the unconditional maximum likelihood method. Instead of taking the sensitivity analysis approach, Huang et al.\cite{huang2023} successfully introduced the inverse probability weighting method (IPW) to adjust the publication bias based on $t$-type selection functions. This method is attractive since it utilizes more interpretable selection functions based on the $t$-type statistics and enables us to obtain a closed-form estimate without relying on the sensitivity analysis approach. This paper aimed to extend their work to a more complicated case of NMA. 

We start with preparing the notations of extended data structure using clinical trial registries. Let $M_k$ denote the number of unpublished studies in the study design $k$, and $M=\sum_{k=1}^K M_k$ denote the total number of unpublished studies in an NMA. We assume all of them could be identified by a comprehensive search from multiple clinical trial registries. Then, for each study design $k$, we have $S_k=N_k+M_k$ studies. In total, we identified $S=\sum_{k=1}^K S_k$ studies in an NMA. Let $Z_{ik}$ indicate the publication status of the $i$th study in the design $k$; $Z_{ik}=1$ if published and $Z_{ik}=0$ otherwise. For the $N$ published studies ($Z_{ik}=1$), we assume the vector ($\mathbf{y_{ik},\Sigma_{ik},n_{ik}}$) is available. For the $M$ unpublished studies ($Z_{ik}=0$) from clinical trial registries, we assume the planned sample size $n_{ik}$ is available and it is consistent with the actual sample size of those unpublished studies. We assume these $N+M$ studies represent a random sample from the study population of interest.

\subsection{Selection functions}
Selection functions are used to describe how the probability of a study being published. Following the previous work by Huang et al.\citep{huang2023}, we proposed several $t$-type selection functions to model the selective publication process in NMAs. Let $\pi_{ik}=P(Z_{ik}=1\mid \mathbf{y_{ik}},\mathbf{\Sigma_{ik}},n_{ik})$ denote the publishing probability of the $i$th study in design $k$ given its outcomes measures $(\mathbf{y_{ik}},\mathbf{\Sigma_{ik}},n_{ik})$. Let $t_{ik}^*$ be a $t$-type statistic, which would be a key statistic for scientific argument in the $i$th study of the $k$th design type. For two-arm studies ($k=1,2$ and $3$), it is the $t$-statistic for the two-arm comparison. For the multi-arm studies, we have multiple $t$-statistics. We carefully choose one of them, which is more likely to be responsible for publication. One possible way is to follow the approach by Noortgate et al.\citep{van2015}; we assume the selective publication process only depends on the most remarkable findings, that is $t_{i4}=max(y_{i4}^{AC}/s_{ik}^{AC},y_{i4}^{BC}/s_{ik}^{BC})$ if higher $t$-statistic is preferred in practice, and $t_{i4}=min(y_{i4}^{AC}/s_{ik}^{AC},y_{i4}^{BC}/s_{ik}^{BC})$ vice versa. We consider three types of selection functions based on $t_{ik}^*$ as follows:
\begin{itemize}
    \item {Common-intercept selection functions:} As the model of less number of unknown parameters, one may consider a logistic-type selection function with the common intercept and slope across the design types, 
\begin{equation}
 \pi_{ik}(\beta_0,\beta_1) = 
\dfrac{\exp{(\beta_0+\beta_1 t_{ik}^*)}}
{1+\exp{(\beta_0+\beta_1 t_{ik}^*)}},
\label{logistic2}
\end{equation}
or a probit type,
\begin{equation}
 \pi_{ik}(\beta_0,\beta_1) = \Phi(\beta_0+\beta_1 t_{ik}^*),
\label{probit2}
\end{equation}
\end{itemize}
where $\Phi(\cdot)$ is the cumulative distribution function of the standard normal distribution. In these selection functions, the intercept and slope parameters are common across the design types and then, the selective publication probability is the same over the design types given the $t$-statistics.  
\begin{itemize}
\item {Design-specific intercept selection functions:} The common intercept and slope assumption in ($\ref{logistic2}$) and ($\ref{probit2}$) may be too restrictive to flexibly describe actual selective publication processes. Then, one may allow the design-specific intercepts in the selection functions. The design-specific intercept logistic selection function is defined as
\begin{equation}
 \pi_{ik}( {\beta_{0k}},\beta_1) = 
\dfrac{\exp{({\beta_{0k}}+\beta_1t_{ik}^*)}}
{1+\exp{({\beta_{0k}}+\beta_1t_{ik}^*)}},
\label{logistic5}
\end{equation}
and the design-specific intercept probit selection function is defined as

\begin{equation}
 \pi_{ik}( {\beta_{0k}},\beta_1) = \Phi({\beta_{0k}}+\beta_1 t_{ik}^*).
\label{probit5}
\end{equation}
\end{itemize}

\begin{itemize}
\item {Separate selection functions:} As the most general form of the selective publication process, one may consider separate selection functions over the design types. The separate logistic selection function is defined as
\begin{equation}
 \pi_{ik}( {\beta_{0k},\beta_{1k}}) = 
\dfrac{\exp{({\beta_{0k}}+{\beta_{1k}}t_{ik}^*)}}
{1+\exp{({\beta_{0k}}+{\beta_{1k}}t_{ik}^*)}},
\label{logistic8}
\end{equation}
and the probit one is as
\begin{equation}
 \pi_{ik}( {\beta_{0k},\beta_{1k}}) = \Phi({\beta_{0k}}+{\beta_{1k}}t_{ik}^*).
\label{probit8}
\end{equation}
\end{itemize}

  Among the three types of selection functions, the separate selection function is the most general and the other two are special cases of it. Thus, it is desirable to apply the separate selection function. However, in practice, it is often the case that some design types have a limited number of studies. In such cases, we need to consider more restrictive models, which are stably estimable. 

\subsection{Inference procedure via the inverse probability weighting }
To estimate the unknown parameters in the selection functions, we use the similar estimating equations proposed by Huang et al.\citep{huang2023}. By incorporating the information from clinical trial registries, we consider the estimating equations in the form of 
\begin{equation}
U(\mathbf{\beta})=\sum_{k=1}^K\sum_{i=1}^{S_k}\left\{1-\frac{Z_{ik}}{\pi_{ik}(\mathbf{\beta})}\right\}g(n_{ik})=0,
\label{eq0}
\end{equation}
where $\mathbf{\beta}$ is a vector of unknown parameters in the selection function and $g(n_{ik})$ is a function of the same dimension as $\mathbf \beta$. This estimating equation is used in the missing data problem and causal inference area for the propensity score analysis under the missing not at random settings\citep{kott2010,miao2016,morikawa2021}. For the common-intercept selection functions such as (\ref{logistic2}) and (\ref{probit2}), we use the following estimating equations
\begin{equation}
U(\beta_0,\beta_1)=\sum_{k=1}^K\sum_{i=1}^{S_k}\left\{1-\dfrac{Z_{ik}}{\pi_{ik}(\beta_0,\beta_1)}\right\} 
\left(
 \begin{array}{c}
      1 \\
      \sqrt{n_{ik}}\\
      \end{array}
\right)=0.
\label{eq1}
\end{equation}
For the design-specific selection functions such as (\ref{logistic5}) and (\ref{probit5}), we must prepare for a five-dimensional vector for $g(n_{ik})$. For example, we can consider the following estimating equation; 

\begin{equation}
U({\beta_{01}},\beta_{02},\cdots,\beta_{0K},\beta_1)=\sum_{k=1}^K\sum_{i=1}^{S_k}\left\{1-\dfrac{Z_{ik}}{\pi_{ik}({\beta_{0k}},\beta_1)}\right\} 
\left(
 \begin{array}{c}
      1 \\
      \sqrt{n_{ik}}\\n_{ik}\\n_{ik}^{1.5}\\n_{ik}^2
      \end{array}
\right)=0.
\label{eq2}
\end{equation}

For the separate selection functions such as (\ref{logistic8}) and (\ref{probit8}), we consider the following estimating equations
\begin{equation}
U({\beta_{0k},\beta_{1k}})=\sum_{i=1}^{S_k}\left\{1-\dfrac{Z_{ik}}{\pi_{ik}( {\beta_{0k},\beta_{1k}})}\right\} 
\left(
 \begin{array}{c}
      1 \\
      \sqrt{n_{ik}}\\
      \end{array}
\right)=0.
\label{eq3}
\end{equation}

Let the solution to the estimating equation ($\ref{eq0}$) denoted by $\mathbf{\hat{\beta}}$ and $\hat\pi_{ik}=\pi_{ik}(\mathbf{\hat{\beta}})$ be the resulting publishing probability for the $i$th study in design $k$. With similar arguments to Huang et al., one can show that $\mathbf{\hat{\beta}}$ is a consistent estimator of $\mathbf{\beta}$ if the selection function is correctly specified (for details see Section 3.3 of Huang et al.\citep{huang2023}). 

To estimate the overall treatment effects $\mathbf{\mu_k}$ as well as the design-specific between-study heterogeneity $\mathbf{\tau_k}$, Huang et al.\cite{huang2023} proposed an IPW-based simple estimator of a closed-form expression for the population mean, as well as that for the between-study variance. However, it seems impossible to obtain simple closed-form estimators for the NMA. Instead, we propose to estimate the vector ($\mathbf{\mu_k},\mathbf{\tau_k}$) by maximizing the IPW version of the log-likelihood function,

\begin{align}
    logL_{IPW}(\mathbf{\mu,\tau})=-\dfrac{1}{2}\sum_{k=1}^K\sum_{i=1}^{N_k} \dfrac{Z_{ik}}{\hat\pi_{ik}}\left\{log\mid \mathbf{\Omega_{ik}+\Sigma_{ik}}\mid+(\mathbf{y_{ik}-\mu_k})^T(\mathbf{\Omega_{ik}+\Sigma_{ik}})^{-1}(\mathbf{y_{ik}-\mu_k})\right\}
\label{log.ipw}
\end{align}
The maximizer is denoted by $(\hat\mu_k^{IPW},\hat\tau_k^{IPW})$, which is an IPW version of the maximum likelihood estimator. For the construction of confidence interval, Huang et al.\cite{huang2023} reported that for the pairwise MA, the asymptotic variance might be unsatisfactory due to the nearly singular hessian matrix in some cases, and then parametric bootstrap confidence intervals were recommended for practical use. In this paper,
we adopt the same idea to construct the confidence intervals for $(\hat\mu_k^{IPW},\hat\tau_k^{IPW})$ with the parametric bootstrap approach. First, we generate the parametric bootstrap samples $\mathbf{\tilde{y}}_{ik}$ from $N(\hat\mu_k^{IPW},\hat{\Omega}_{ik}+\Sigma_{ik})$. Then we solve the following estimating equation 
\begin{equation}
U(\mathbf{\tilde{\beta}})=\sum_k^K\sum_{i=1}^{S_k}\left\{1-\frac{Z_{ik}}{{\pi}_{ik}(\mathbf{\tilde{\beta}})}\right\}g(n_{ik})=0,
\label{eq0.boot}
\end{equation}
where ${\pi}_{ik}(\mathbf{\tilde{\beta}})$ is defined as publishing probability of bootstrap samples, which can be obtained by replacing the $\mathbf{y}_{ik}$ with the bootstrapped $\mathbf{\tilde{y}}_{ik}$ in $\pi_{ik}(\mathbf{\beta})$. By plugging in ${\pi}_{ik}(\mathbf{\tilde{\beta}})$ and replacing the $\mathbf{y}_{ik}$ with the bootstrapped $\mathbf{\tilde{y}}_{ik}$ in log-likelihood ($\ref{log.ipw}$), we define the corresponding solution as $(\tilde\mu_k^{IPW},\tilde\tau_k^{IPW})$ for the bootstrap samples. Repeating this procedure for $B$(say, 1000) bootstrap samples,  
we could construct the bootstrap sampling distribution to approximate the real distribution of the parameters. Let $\tilde\mu_k^{IPW(b)}$ denote the estimates with the $b$th bootstrap sample. We define the bootstrap standard deviation as $\sigma_k^{boot}=\sqrt{B^{-1} \sum_{b=1}^B (\tilde\mu_k^{IPW(b)}-\bar{\mu}_k^{boot})}$, where $\bar{\mu}_k^{boot}=B^{-1} \sum_{b=1}^B \tilde\mu_k^{IPW(b)}$ and a bootstrap two-tailed 95 percent confidence interval is constructed by ($\hat{\mu}_k^{IPW}+q(0.025) \sigma_k^{boot}, \hat{\mu}_k^{IPW}+q(0.975) \sigma_k^{boot}$), where $q(0.025)$ and $q(0.975)$ are the 2.5 and 97.5 percentiles of the standardized bootstrap samples of $ (\tilde\mu_k^{IPW(b)}-\bar{\mu}_k^{boot})/\sigma_k^{boot}$ (see Theorem 23.5 of Van der Vaart \cite{van2000}). The bootstrap confidence interval for $\hat\tau_k^{IPW}$ could be constructed in the same manner.

Besides, we also propose the publication bias-adjusted version of the $P$-score for the treatment ranking. By replacing quantities in the definition of the $P$-score, we define the IPW version of the $P$-score as follows;

      \begin{eqnarray}
        P_{ij}^{IPW} & = & \left\{\begin{array}{ll} \Phi\Biggl(\dfrac{\hat{\mu}_{ij}^{IPW}}{\sigma_{ij}^{boot}}\Biggl)& \mbox{if the benefit is supposed to be positive} \\ 1-\Phi\Biggl(\dfrac{\hat{\mu}_{ij}^{IPW}}{\sigma_{ij}^{boot}}\Biggl) & \mbox{if the benefit is supposed to be negative}\end{array}\right.\mbox{,} \label{pscore.adj}
      \end{eqnarray}
 where $\hat{\mu}_{ij}^{IPW}$ is the estimated treatment effect difference between treatment $i$ and $j$ based on the IPW method, and  $\sigma_{ij}^{boot}$ is the corresponding standard error from bootstrap samples. Then the adjusted $P$-score could be calculated based on the $P_{ij}^{IPW}$ in the same way as the original $P$-score.

\section{Simulation study}
We conducted a comprehensive simulation study to evaluate the performance of our proposed methods in the following aspects: a) the performance of our proposed IPW estimator when the selection function is correctly specified and misspecified; b) the influence of the assumption for the heterogeneity structure (common heterogeneity versus design-specific heterogeneity); c) the finite-sample performance of the proposed methods when we only have very few studies for each study design.

\subsection{Data generation}
We first introduce how to generate the complete data of a network meta-analysis using the same framework in Section 2.2. Suppose we are doing an NMA with dichotomous data, and the log-odds ratio was used as the summary measure to compare the efficacy of two active treatments $A$ and $B$ in contrast to the control group $C$ (T=3). Four types of study designs were considered: $T_1=\left\{A,C\right\}$,\ $T_2=\left\{B,C\right\}$,\ \ $T_3=\left\{A,B\right\}$,\ and\ $T_4=\left\{A,B,C\right\}$.
Motivated by the example in Section 5, we set $\mu^{AC}=0.5$ and $\mu^{BC}=0.3$ in Models (1) and (2), which corresponds to the efficacy of Paroxetine and Fluvoxamine in contrast to Placebo, respectively. We assume that consistency holds for $\mu^{AB}=\mu^{AC}-\mu^{BC}$. Two types of heterogeneity structures were considered; one is the common heterogeneity structure that $\tau_{1}^2=\tau_{2}^2=\tau_{3}^2=\tau_{4}^2=$ 0.0025 to 0.09, and the other is design-specific heterogeneity structure that $\tau_{1}^2=0.0025$, $\tau_{2}^2=0.0225$, $\tau_{3}^2=0.04$, $\tau_{4}^2=0.09$, which reflected the small to moderate heterogeneity in practice. We considered three settings on the number of studies in each design type. In the moderate-size and large-size settings, $S_k$ was set as 10 and 20, respectively, for all the design types. We also considered the small-size case, in which $S_k$ was generated from the discrete uniform distribution on $U$(5,10).  Similar to Huang et al.\cite{huang2023}, we begin with generating the individual patient data. In this network MA, we have two-arm trial designs (Model (1)) as well as the three-arm trial design (Model (2)). We generate all the data from the bivariate framework as Model (2) and keep the treatment arms according to the design of consideration. 
For example, we generate S studies from Model (2), and for S1 studies with the design type $k=1$ or $T_1=\left\{A,C\right\}$, the group B is deleted. Similar methods are applied to S2 and S3 studies with the design type $k=2$ and $k=3$, respectively. At first, we generate the true log-odds ratios of $\mu_{ik}^{AC}$ and $\mu_{ik}^{BC}$ in the $i$th study of design $k$ from $N(\mu^{AC},\tau_k^2)$ and $N(\mu^{BC},\tau_k^2)$, respectively, and we generate the true event rate of the control group \emph{$p_{ik}^C$} in the same study from the uniform distribution of \emph{$U$ (0.2,0.9)}. Then the event rate in the treatment group A (\emph{$p_{ik}^A$}) could be derived as $e^{\mu_{ik}^{AC}}p_{ik}^C/(1-p_{ik}^C+p_{ik}^Ce^{\mu_{ik}^{AC}})$, and the event rate in the treatment group B (\emph{$p_{ik}^B$}) could be derived in the same way. We considered the balanced trial design, therefore, the number of subjects in the three groups is assumed identical. Following Kuss\cite{kuss2015}, the total sample size for each comparison (AC and BC) was generated from \emph{LN(5,1)}, the log-normal distribution with the location parameter 5 and scale parameter 1, and the minimum sample size was restricted to 20 patients (values below 20 were rounded up to 20). Then, the individual participant data could be obtained by generating the responders from the binomial distributions \emph{B($n_{ik}^{A},p_{ik}^{A}$)}, \emph{B($n_{ik}^{B},p_{ik}^{B}$)} and \emph{B($n_{ik}^{C},p_{ik}^{C}$)}, where $n_{ik}^{A}=n_{ik}^{B}=n_{ik}^{C}$. With the generated individual participant data, we calculated the empirical log-odds ratios and the corresponding variance-covariance matrix in Model (2). For the studies with the design type $d=4$, we need to have empirical covariance in Model (2). It is calculated by $cov(y_{i4}^{AC},y_{i4}^{BC})=var[logit(P_{i4}^C)]$, since $var(y_{i4}^{AC})=var([ogit(P_{i4}^A)]+var[logit(P_{i4}^C)]$ and $var(y_{i4}^{BC})=var[logit(P_{i4}^B)]+var[logit(P_{i4}^C)]$.

Next, we illustrate how we selectively pick up studies to be unpublished from the complete data. The indicator of publication status \emph{$Z_{ik}$} was generated from the binomial distribution \emph{$B(1,\pi_{ik}(\mathbf\beta))$}, where $\pi_{ik}(\mathbf\beta)$ is the corresponding publishing probability for the $i$th study in design $k$ conditional on its empirical $t$ statistic for the log odds-ratio. We considered two types of selection functions for the $\pi_{ik}(\mathbf\beta)$; the common-intercept selection function of the logistic one ($\ref{logistic2}$) with $\beta_0=-0.2$ and $\beta_1=0.8$, and the design-specific intercept selection function of the logistic one ($\ref{logistic5}$) with $\mathbf{\beta}_{0k}=\{-0.3, 0.4, 0.3, 0.2\}$ and $\beta_1=1$. Both of these two selection functions can result in around 30\% unpublished studies. 

\subsection{Applied methods}
For comparison, we implemented both the standard multivariate random-effects model (MRE) in Section 2 and the proposed IPW-based publication bias adjustment method (IPW) in Section 3 to show the impacts with and without accounting for the selective publication process. $nlminb()$ function in R software was utilized to optimize the log-likelihoods of (\ref{mle}) and (\ref{log.ipw}). The performance of both methods was assessed by the mean and standard deviation of the estimates for $\mu^{AC}$ and $\mu^{BC}$ as well as the average lengths and coverage probabilities of the confidence intervals (we generate 1000 bootstrap samples to construct the parametric bootstrapping CIs which was introduced in Section 3.3). For the estimates of design-specific heterogeneity $\tau_d$, instead of average lengths of confidence intervals, we count the numbers of zero estimates.

\subsection{Results}
In Table 1, we present the simulation results for the estimation of $\mu^{AC}$ and $\mu^{BC}$ when the studies were selected to be published under the design-specific intercept logistic selection function ($\ref{logistic5}$) for the moderate- and the large-size cases (results for small-size cases are presented in Web-Appendix). Since we focus on the case of $d=4$ according to the setting in Section 2.2, we applied six types of selection functions. The common-intercept logistic selection function ($\ref{logistic2}$) has 2 unknown parameters. Then, we call this model the "2-logit". Similarly, the design-specific intercept and the separate logistic selection functions ($\ref{logistic5}$) and ($\ref{logistic8}$) are referred to as "5-logit" and "8-logit", respectively. Similar terminologies are used for the probit-type selection models. Since the datasets were generated under the "5-logit" model, the "5-logit" and "8-logit" models were correctly specified. Without considering the publication bias, the standard multivariate random-effects model implemented with the published studies substantially overestimates both treatment effects of $\mu^{AC}$ and $\mu^{BC}$. The resulting coverage probabilities were far beyond the nominal level, especially for $\mu^{AC}$ estimates, which suggested that our simulation successfully generated NMAs on which a selective publication process certainly influenced. On the other hand, all of our proposed IPW estimators gave less biased estimates. When the selection function was correctly specified, which referred to the results using "5-logit" and "8-logit, estimates showed smaller bias in most scenarios, while the coverage probabilities might be below the nominal level when the between study heterogeneity is moderate ($\tau_k=0.3$). On the other hand, when the selection function was misspecified, the coverage probabilities could still be highly improved in contrast to the multivariate random-effects model. Especially the estimates using misspecified "2-logit" which showed satisfactory performance in the perspectives of better coverage probabilities and shorter confidence intervals.

Next, we present the simulation results for the heterogeneity estimates when $S_k$ equals 10 in Table 2 (for the results when $S_k$ equals 20, see Web Appendix B-2). Overall, our IPW version heterogeneity estimators had smaller biases and fewer zero estimates in contrast to the estimators from the standard multivariate random-effects model without accounting for the publication bias. Similar to the observations in the simulation by Huang et al.\cite{huang2023}, the misspecification of the selection functions seemed to have no big influence on the heterogeneity estimates. However, we observed that no matter whether the data was generated under the common heterogeneity or design-specific heterogeneity structure, the estimates for the heterogeneity of direct evidence ($\tau_1$, $\tau_2$) seem poorer than the estimates from the indirect evidence ($\tau_3$) and multi-arm trials ($\tau_4$). A similar trend also appeared in the standard multivariate random-effects meta-analysis.

In the Web Appendix, we present the additional simulation results for the common-intercept selection function (see Web Appendix A) and small-size cases (see Web Appendix A-3 and Web Appendix B-3). Here, we briefly summarized the key findings. When the studies were selected with the two-parameter logistic function, we observed larger bias in the $\mu^{AC}$ and $\mu^{BC}$ estimates from the IPW estimators using misspecified selection functions, especially the IPW estimators using the three types of probit selection functions.
Although the coverage probabilities remained the same as we presented in Table 1. In the small-size NMA case, we found that IPW estimator using a two-parameter logistic selection function outperforms other estimators in general with less bias and higher coverage probability, even when the selection function was misspecified. This phenomenon indicates that when only a few studies are available in a network meta-analysis, the simplest selection function with fewer parameters might give better performance.

\section{Data application}
Cipriani et al. \citep{cipriani2018} conducted a network meta-analysis to compare the efficacy of 21 antidepressant drugs for the acute treatment of adults with unipolar major depressive disorder. For convenience of illustration, here we focus on the efficacy of the two most commonly utilized drugs (Fluvoxamine and  Paroxetine) in contrast to a control group of placebo. The outcome was defined as the response rate of patients who had a
reduction of $\ge50\%$ of the total score on a standardized observer-rating scale for depression and measured as a log-odds ratio. Hence, the higher response rate indicates that potentially patients will benefit more from this treatment than the other. When considering the potential selective publication process, $t$-statistics with larger positive values might be more likely to be published ($t_{ik}^*=max(\cdot)$). To screen all the eligible studies, Cipriani et al. \citep{cipriani2018} did a comprehensive search across different databases as well as the clinical trial registries, both published and unpublished studies were identified which involved 97 studies. In Table 3, we show the detailed publication status of this NMA concerning different study designs. In total, we have 28 ( around 30\% ) unpublished studies. With their efforts to collect the outcomes of as many studies as possible, Cipriani et al. \citep{cipriani2018} obtained the outcomes of 22 studies among 28 unpublished studies. For the remaining 6 studies, only the information on sample sizes is available. In real meta-analyses, one may try to identify available outcomes as comprehensively as possible and may use the outcomes of the 22 unpublished studies. For illustration, in applying our proposed method and the standard multivariate random-effect model, we regard the 69 published studies as published and the 28 unpublished studies as unpublished. To the published 69 studies, we applied the standard multivariate random-effects model (2) for the NMA. In addition, we applied this model to the 81 studies of the outcome (69 published and 22 unpublished) for reference. Although we missed the outcomes of 6 studies, it would provide an estimate less influenced by the selective publication process and it is regarded as a gold standard estimate in interpreting all the analysis results.  To present the information of published studies as well as those unpublished,
we propose to use the modified comparison-adjusted funnel plot (see Figure 1). As Huang et al.\cite{huang2021} did in their paper, we simply added horizontal lines passing by the y-axis at $\sqrt{n_{ik}}$  in the scatter plot to represent the additional information of sample sizes of unpublished studies and different colors were utilized to distinguish the different study designs. We applied Egger's regression test to the comparison-adjusted effect sizes of the published studies, which suggested the funnel-plot asymmetry with a $P$-value of 0.025. 

In Table 4, we summarized estimates for the log-odds ratios among the three treatments by the proposed methods with several selection functions. By utilizing the published 69 studies, we had the estimate of the comparison of Paroxetine versus Placebo as 0.594 ( 95\% CI: [0.496,0.693] ) and that of Fluvoxamine versus Placebo as 0.411 ( 95\% CI: [0.253,0.569] ). On the other hand, the gold standard estimates of the log-odds ratio with the additional 22 studies were 0.503 ( 95\% CI: [0.413,0.593] ) for the comparison of Paroxetine versus Placebo and 0.348 ( 95\% CI: [0.215,0.481] ) for the comparison of Fluvoxamine versus Placebo. From these two analyses, we observed around 18\% difference for both comparisons (Paroxetine versus Placebo and Fluvoxamine versus Placebo), which suggests that estimates based on the 69 published studies suffered from selective publication. On the other hand, we observed that our methods can substantially reduce the overestimated treatment effects. For the efficacy of Paroxetine versus Placebo, our methods yielded estimates from 0.488 to 0.535, and for that of Fluvoxamine versus Placebo, our estimates ranged from 0.213 to 0.307. Among these results, we found that the complete design-specific selection function of 8-parameter logistic one showed very similar results with gold standard estimates of 0.518 (95\% CI: [0.401,0.633]) for the efficacy of Paroxetine versus Placebo, and 0.307 (95\% CI: [0.127,0.491]) for the efficacy of Fluvoxamine versus Placebo. 

To further compare the benefits of these two treatments, we also presented the ranking results in Table 6. We observed that all the methods confirmed that Paroxetine has the best efficacy in contrast to Fluvoxamine and Placebo, then is the Fluvoxamine, even without considering the publication bias. This is probably due to the big difference in the efficacy among these treatments; selective publication did not have an impact enough to switch the ranking. The $P$-scores based on 69 published and 22 unpublished studies were estimated as 0.991 for the Paroxetine and 0.509 for the Fluvoxamine. Similar to our observation in Table 2, the adjusted $P$-scores using the flexible 8-parameter selection functions gave the very closed estimates (0.993 (8-parameter logistic) and 0.990 (8-parameter probit)  for the Paroxetine; 0.507 (8-parameter logistic) and 0.508 (8-parameter probit) for the Fluvoxamine). Hence, when the selection function successfully captured the potential selection process, our proposed adjusted $P$-scores could also properly rank the treatments.

As a secondary interest, we also present the estimates of the design-specific heterogeneity in Table 5. We observed that the gold-standard estimates for heterogeneity gave non-zero estimations in three study designs. Among them, studies about Fluvoxamine versus Placebo are the most heterogeneous ones ($\tau_2$=0.334), then are the studies with three-arm designs ($\tau_4$=0.297) and the studies about Paroxetine versus Placebo ($\tau_2$=0.086). This may indicate that a common-heterogeneity structure might not be appropriate in this data. By accounting for the publication bias, our proposed methods all gave the non-zero estimate to the heterogeneity of studies about Fluvoxamine versus Placebo ($\tau_2$=0.421 to 0.483). Although it is much larger than the estimate only with 69 published studies ($\tau_2$=0.382), which also only yield a non-zero estimate in this study design.

\section{Discussion}
Nowadays, many clinical trial registries have been well-established and are suggested as a very important procedure for searching studies in Cochrane reviews\cite{higgins2023}. However, the huge information in the clinical trial registries has not been fully utilized, particularly in the statistical analysis for publication bias. Recently, the authors have pointed out that information in clinical trial registries makes the data structure of meta-analysis look like the datasets arising in general missing data problem and there are potentials to handle the publication bias issue like the standard missing data problem. Indeed, in a series of their work\cite{mavridis2013,huang2021,huang2023}, they successfully introduced some typical missing data analysis methods to the publication bias issue.
Huang et al.\cite{huang2023} introduced a simple IPW estimator of a closed form to adjust the publication bias in the pairwise MA. This paper is an extension of the work by Huang et al. to the NMA. Although the proposed estimator does not have a simple closed-form expression, the IPW maximum likelihood is successfully applied to the NMA, and maximizing the IPW log-likelihood is shown to be tractable in our numerical studies. To our best knowledge, this is the first proposal to handle the $t$-type selection functions in the NMA. On practical issues in the utilization of clinical trial registries, one could refer to the detailed discussion by Huang et al.\cite{huang2023} (see also Web-Appendix F in the supporting information).

 Here, we mention some limitations in our model. First, to synthesize the data from different study designs, we assumed the consistency assumption holds in our analysis. Various methods have been established to assess the consistency between the direct and indirect evidence\cite{donegan2013}. We emphasize that one should carefully evaluate the characteristics of the included studies (e.g., participants, dose of interventions), and a possible solution could be introducing the inconsistency as a special type of heterogeneity in the standard model\cite{higgins2012}. The potential impacts of the selective publication on the validity of consistency assumption have not been addressed so far. We would like to address this issue in our future work.

It is well-known that network meta-analysis could be fitted with either contrast-based or arm-based hierarchical model\cite{white2019}. In the example and the simulation studies in this paper, we considered the network MA of dichotomous outcomes. A contrast-based model was utilized with the summary measure of the log-odds ratio between the two treatment groups, and the corresponding likelihood was modeled with normal approximation. As an alternative, one could also model the number of events in each treatment group with the binomial likelihood or model the log odds of the events with the normal likelihood. The difference between these two types of models has already been discussed a lot ( for a detailed discussion see Sanalti et al.\cite{salanti2008} and White et al.\cite{white2019}), and we acknowledged that our IPW framework also has potentials to be extended to these arm-based models. A similar adoption of our methodology to these alternative models could be weighting the likelihood of each published study with the estimated publishing probability under the different selection processes we introduced in Section 3.2.

Finally, we believe that the utilization of information from clinical trial registries deserves more attention in addressing the publication bias issue in the general meta-analysis field (e.g., pairwise MA, NMA). Similar to the statement in the latest version (6.4) of the Cochrane Handbook for Systematic Reviews of Interventions, which stratifies the values of identifying unpublished studies, since the impact of selective publication can be quantified more easily with a known number of unpublished studies\cite{higgins2023}. Here, we want to further emphasize that it is also very important to consider how to utilize the limited information of those identified but unpublished studies at hand. Appropriate statistical models that can incorporate this information (e.g., planned sample size) to make robust inferences should be developed. Our method is to this end.

\bibliography{wileyNJD-AMA}

\begin{thebibliography}{10}
\providecommand \doibase [0]{http://dx.doi.org/}%

\bibitem{rouse2017}
Rouse B, Chaimani A, Li T. Network meta-analysis: an introduction for
  clinicians. {\it Internal and emergency medicine} 2017\string; 12(1)\string:
  103--111.

\bibitem{salanti2011}
Salanti G, Ades A, Ioannidis JP. Graphical methods and numerical summaries for
  presenting results from multiple-treatment meta-analysis: an overview and
  tutorial. {\it Journal of clinical epidemiology} 2011\string; 64(2)\string:
  163--171.

\bibitem{efthimiou2016}
Efthimiou O, Debray TP, Valkenhoef vG, et al. GetReal in network meta-analysis:
  a review of the methodology. {\it Research synthesis methods} 2016\string;
  7(3)\string: 236--263.

\bibitem{antoniou2023}
Antoniou SA, Mavridis D, Tsokani S, Morales-Conde S, Thanjakumar
  EGSOMVNCFMPSDA. Network meta-analysis as a tool in clinical practice
  guidelines. {\it Surgical Endoscopy} 2023\string; 37(1)\string: 1--4.

\bibitem{balduzzi2023}
Balduzzi S, R{\"u}cker G, Nikolakopoulou A, et al. netmeta: An R package for
  network meta-analysis using frequentist methods. {\it Journal of Statistical
  Software} 2023\string; 106\string: 1--40.

\bibitem{jin2015}
Jin ZC, Zhou XH, He J. Statistical methods for dealing with publication bias in
  meta-analysis. {\it Statistics in medicine} 2015\string; 34(2)\string:
  343--360.

\bibitem{cha2013}
Chaimani A, Higgins JP, Mavridis D, Spyridonos P, Salanti G. Graphical tools
  for network meta-analysis in STATA. {\it PloS one} 2013\string; 8(10)\string:
  e76654.

\bibitem{egger1997}
Egger M, Smith GD, Schneider M, Minder C. Bias in meta-analysis detected by a
  simple, graphical test. {\it BMJ} 1997\string; 315(7109)\string: 629--634.

\bibitem{begg1994}
Begg CB, Mazumdar M. Operating characteristics of a rank correlation test for
  publication bias. {\it Biometrics} 1994\string: 1088--1101.

\bibitem{marks2020}
Marks-Anglin A, Chen Y. A historical review of publication bias. {\it Research
  synthesis methods} 2020\string; 11(6)\string: 725--742.

\bibitem{copas2000}
Copas J, Shi JQ. Meta-analysis, funnel plots and sensitivity analysis. {\it
  Biostatistics} 2000\string; 1(3)\string: 247--262.

\bibitem{copas2001}
Copas J, Shi JQ. A sensitivity analysis for publication bias in systematic
  reviews. {\it Statistical Methods in Medical Research} 2001\string;
  10(4)\string: 251--265.

\bibitem{heckman1979}
Heckman JJ. Sample selection bias as a specification error. {\it Econometrica}
  1979\string; 47(1)\string: 153--61.

\bibitem{preston2004}
Preston C, Ashby D, Smyth R. Adjusting for publication bias: modelling the
  selection process. {\it Journal of Evaluation in Clinical Practice}
  2004\string; 10(2)\string: 313--322.

\bibitem{copas2013}
Copas JB. A likelihood-based sensitivity analysis for publication bias in
  meta-analysis. {\it Journal of the Royal Statistical Society: Series C
  (Applied Statistics)} 2013\string; 62(1)\string: 47--66.

\bibitem{moreno2009}
Moreno SG, Sutton AJ, Ades A, et al. Assessment of regression-based methods to
  adjust for publication bias through a comprehensive simulation study. {\it
  BMC medical research methodology} 2009\string; 9\string: 1--17.

\bibitem{marks2022}
Marks-Anglin A, Luo C, Piao J, et al. EMBRACE: An EM-based bias reduction
  approach through Copas-model estimation for quantifying the evidence of
  selective publishing in network meta-analysis. {\it Biometrics} 2022\string;
  78(2)\string: 754--765.

\bibitem{mavridis2013}
Mavridis D, Sutton A, Cipriani A, Salanti G. A fully Bayesian application of
  the Copas selection model for publication bias extended to network
  meta-analysis. {\it Statistics in medicine} 2013\string; 32(1)\string:
  51--66.

\bibitem{mavridis2014selection}
Mavridis D, Welton NJ, Sutton A, Salanti G. A selection model for accounting
  for publication bias in a full network meta-analysis. {\it Statistics in
  medicine} 2014\string; 33(30)\string: 5399--5412.

\bibitem{huang2023}
Huang A, Morikawa K, Friede T, Hattori S. Adjusting for publication bias in
  meta-analysis via inverse probability weighting using clinical trial
  registries. {\it Biometrics} 2023.

\bibitem{rucker2015}
R{\"u}cker G, Schwarzer G. Ranking treatments in frequentist network
  meta-analysis works without resampling methods. {\it BMC medical research
  methodology} 2015\string; 15(1)\string: 1--9.

\bibitem{nikolakopoulou2021}
Nikolakopoulou A, Mavridis D, Chiocchia V, Papakonstantinou T, Furukawa TA,
  Salanti G. Network meta-analysis results against a fictional treatment of
  average performance: treatment effects and ranking metric. {\it Research
  synthesis methods} 2021\string; 12(2)\string: 161--175.

\bibitem{chaimani2021}
Chaimani A, Porcher R, Sbidian {\'E}, Mavridis D. A Markov chain approach for
  ranking treatments in network meta-analysis. {\it Statistics in medicine}
  2021\string; 40(2)\string: 451--464.

\bibitem{brignardello2018}
Brignardello-Petersen R, Johnston BC, Jadad AR, Tomlinson G. Using decision
  thresholds for ranking treatments in network meta-analysis results in more
  informative rankings. {\it Journal of Clinical Epidemiology} 2018\string;
  98\string: 62--69.

\bibitem{chiocchia2020}
Chiocchia V, Nikolakopoulou A, Papakonstantinou T, Egger M, Salanti G.
  Agreement between ranking metrics in network meta-analysis: an empirical
  study. {\it BMJ open} 2020\string; 10(8)\string: e037744.

\bibitem{deangelis2005}
DeAngelis CD, Drazen JM, Frizelle FA, et al. Clinical trial registration: a
  statement from the International Committee of Medical Journal Editors. {\it
  Archives of Dermatology} 2005\string; 141(1)\string: 76--77.

\bibitem{bian2010}
Bian ZX, Wu TX. Legislation for trial registration and data transparency. {\it
  Trials} 2010\string; 11\string: 1--4.

\bibitem{viergever2015}
Viergever RF, Li K. Trends in global clinical trial registration: an analysis
  of numbers of registered clinical trials in different parts of the world from
  2004 to 2013. {\it BMJ open} 2015\string; 5(9)\string: e008932.

\bibitem{hart2012}
Hart B, Lundh A, Bero L. Effect of reporting bias on meta-analyses of drug
  trials: reanalysis of meta-analyses. {\it BMJ} 2012\string; 344\string:
  d7202.

\bibitem{baudard2017}
Baudard M, Yavchitz A, Ravaud P, Perrodeau E, Boutron I. Impact of searching
  clinical trial registries in systematic reviews of pharmaceutical treatments:
  methodological systematic review and reanalysis of meta-analyses. {\it BMJ}
  2017\string; 356\string: j448.

\bibitem{alqaidoom2022}
Alqaidoom Z, Nguyen PY, Awadh M, Page MJ. Impact of searching clinical trials
  registers in systematic reviews of pharmaceutical and non-pharmaceutical
  interventions: Reanalysis of meta-analyses. {\it Research Synthesis Methods}
  2022.

\bibitem{huang2021}
Huang A, Komukai S, Friede T, Hattori S. Using clinical trial registries to
  inform Copas selection model for publication bias in meta-analysis. {\it
  Research Synthesis Methods} 2021\string; 12(5)\string: 658--673.

\bibitem{van2015}
Noortgate V.~dW, L{\'o}pez-L{\'o}pez JA, Mar{\'\i}n-Mart{\'\i}nez F,
  S{\'a}nchez-Meca J. Meta-analysis of multiple outcomes: A multilevel
  approach. {\it Behavior research methods} 2015\string; 47\string: 1274--1294.

\bibitem{kott2010}
Kott PS, Chang T. Using calibration weighting to adjust for nonignorable unit
  nonresponse. {\it Journal of the American Statistical Association}
  2010\string; 105(491)\string: 1265--1275.

\bibitem{miao2016}
Miao W, Tchetgen~Tchetgen EJ. On varieties of doubly robust estimators under
  missingness not at random with a shadow variable. {\it Biometrika}
  2016\string; 103(2)\string: 475--482.

\bibitem{morikawa2021}
Morikawa K, Kim JK. Semiparametric optimal estimation with nonignorable
  nonresponse data. {\it The Annals of Statistics} 2021\string; 49(5)\string:
  2991--3014.

\bibitem{van2000}
Vaart V.~dAW. {\it Asymptotic statistics}.
\newblock Cambridge University Press .
\newblock 2000.

\bibitem{kuss2015}
Kuss O. Statistical methods for meta-analyses including information from
  studies without any events—add nothing to nothing and succeed nevertheless.
  {\it Statistics in Medicine} 2015\string; 34(7)\string: 1097--1116.

\bibitem{cipriani2018}
Cipriani A, Furukawa TA, Salanti G, et al. Comparative efficacy and
  acceptability of 21 antidepressant drugs for the acute treatment of adults
  with major depressive disorder: a systematic review and network
  meta-analysis. {\it Focus} 2018\string; 16(4)\string: 420--429.

\bibitem{higgins2023}
Higgins~Jpt CJCMLTPMWVe. {\it Cochrane Handbook for Systematic Reviews of
  Interventions version 6.4 (updated August 2023)} .
\newblock 2023.

\bibitem{donegan2013}
Donegan S, Williamson P, D'Alessandro U, Tudur~Smith C. Assessing key
  assumptions of network meta-analysis: a review of methods. {\it Research
  synthesis methods} 2013\string; 4(4)\string: 291--323.

\bibitem{higgins2012}
Higgins JPT, Jackson D, Barrett J, Lu G, Ades A, White I. Consistency and
  inconsistency in network meta-analysis: concepts and models for multi-arm
  studies. {\it Research synthesis methods} 2012\string; 3(2)\string: 98--110.

\bibitem{white2019}
White IR, Turner RM, Karahalios A, Salanti G. A comparison of arm-based and
  contrast-based models for network meta-analysis. {\it Statistics in medicine}
  2019\string; 38(27)\string: 5197--5213.

\bibitem{salanti2008}
Salanti G, Higgins JP, Ades A, Ioannidis JP. Evaluation of networks of
  randomized trials. {\it Statistical methods in medical research} 2008\string;
  17(3)\string: 279--301.

\end{thebibliography}

\newpage
\begin{figure}[t]
\centering\includegraphics[width=18cm, height=10cm]{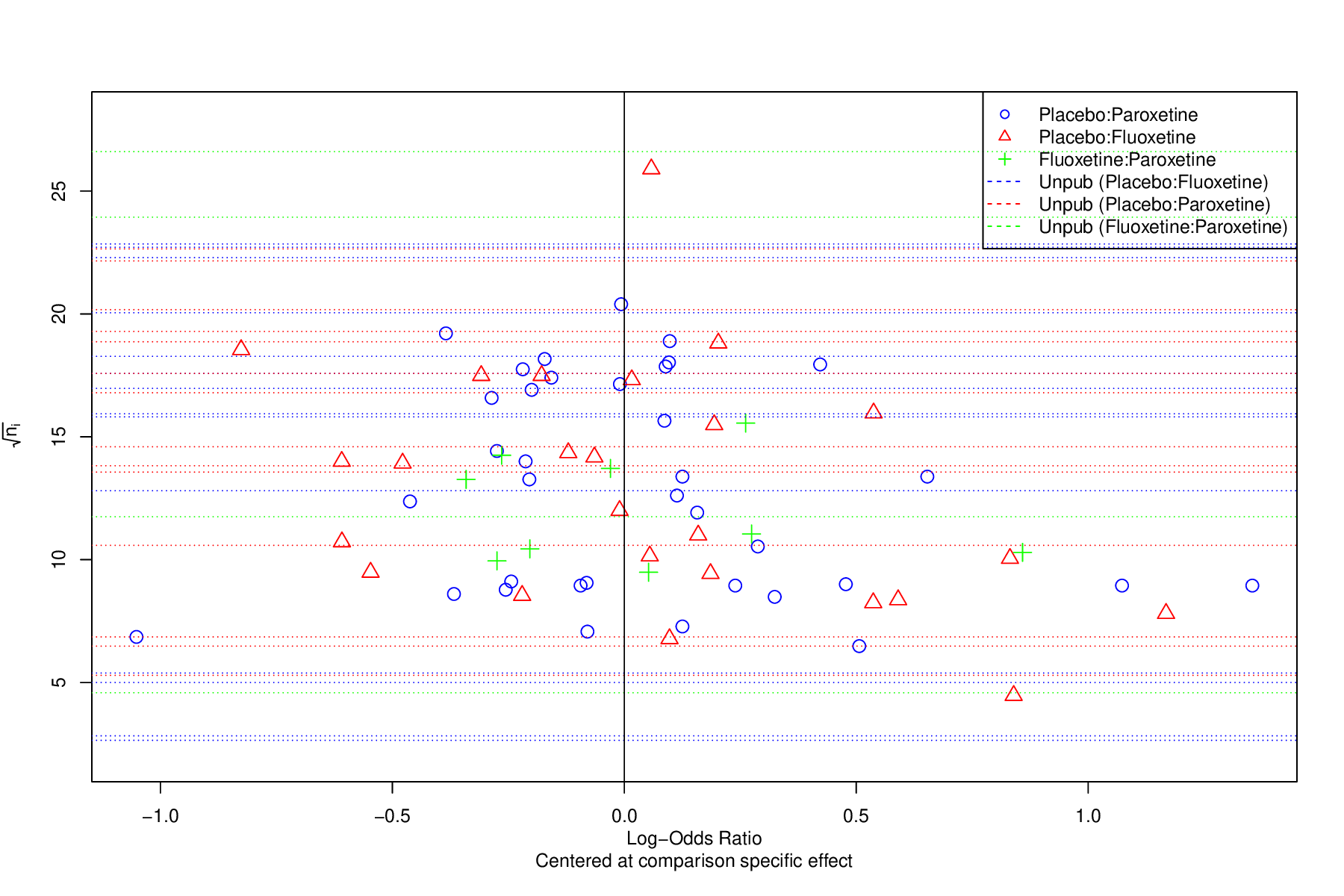}
\caption{\it{Modified comparison-adjusted funnel plot for the network meta-analysis of Antidepressant Drugs}}
\label{fig 4}
\end{figure}

\begin{center}
\begin{table}[h]

\caption{Simulation results for $\mu^{AC}$ and $\mu^{BC}$ under the 5-parameter logistic selection function}
\centering
\resizebox{\linewidth}{!}{
\begin{tabular}[t]{llccccccccccccc}
\toprule
\multicolumn{1}{c}{} & \multicolumn{1}{c}{} & \multicolumn{1}{c}{} & \multicolumn{3}{c}{$\mu^{AC}$ (Moderate-size)} & \multicolumn{3}{c}{$\mu^{BC}$ (Moderate-size)} & \multicolumn{3}{c}{$\mu^{AC}$ (Large-size)} & \multicolumn{3}{c}{$\mu^{BC}$ (Large-size)} \\
\cmidrule(l{3pt}r{3pt}){4-6} \cmidrule(l{3pt}r{3pt}){7-9} \cmidrule(l{3pt}r{3pt}){10-12} \cmidrule(l{3pt}r{3pt}){13-15}
$\tau$ & Method & Selection & AVE(SD) & CP & LOCI & AVE(SD) & CP & LOCI & AVE(SD) & CP & LOCI & AVE(SD) & CP & LOCI\\
\midrule
$\tau_k$ (0.05-0.30) & MRE &  & 0.579 ( 0.088 ) & 0.855 & 0.346 & 0.334 ( 0.1 ) & 0.9 & 0.368 & 0.575 ( 0.063 ) & 0.792 & 0.242 & 0.331 ( 0.068 ) & 0.913 & 0.255\\
& IPW & 2-logit & 0.519 ( 0.088 ) & 0.951 & 0.361 & 0.300 ( 0.106 ) & 0.949 & 0.402 & 0.510 ( 0.062 ) & 0.961 & 0.256 & 0.296 ( 0.077 ) & 0.930 & 0.283\\
 & IPW & 2-probit & 0.488 ( 0.098 ) & 0.964 & 0.418 & 0.288 ( 0.126 ) & 0.936 & 0.467 & 0.477 ( 0.071 ) & 0.974 & 0.312 & 0.282 ( 0.103 ) & 0.903 & 0.340\\
 & IPW & 5-logit & 0.515 ( 0.115 ) & 0.937 & 0.450 & 0.299 ( 0.126 ) & 0.935 & 0.471 & 0.493 ( 0.107 ) & 0.915 & 0.333 & 0.288 ( 0.104 ) & 0.904 & 0.335\\
 & IPW & 5-probit & 0.492 ( 0.099 ) & 0.966 & 0.431 & 0.292 ( 0.130 ) & 0.942 & 0.490 & 0.481 ( 0.069 ) & 0.983 & 0.322 & 0.290 ( 0.107 ) & 0.902 & 0.357\\
 & IPW & 8-logit & 0.537 ( 0.086 ) & 0.940 & 0.350 & 0.303 ( 0.099 ) & 0.939 & 0.357 & 0.520 ( 0.061 ) & 0.941 & 0.238 & 0.298 ( 0.066 ) & 0.932 & 0.240\\
 & IPW & 8-probit & 0.492 ( 0.099 ) & 0.966 & 0.431 & 0.292 ( 0.130 ) & 0.942 & 0.490 & 0.481 ( 0.069 ) & 0.982 & 0.322 & 0.290 ( 0.107 ) & 0.906 & 0.357\\
\addlinespace
common $\tau_k=0.05$ & MRE &  & 0.556 ( 0.068 ) & 0.898 & 0.278 & 0.331 ( 0.072 ) & 0.927 & 0.282 & 0.551 ( 0.049 ) & 0.835 & 0.190 & 0.324 ( 0.048 ) & 0.92 & 0.191\\
 & IPW & 2-logit & 0.513 ( 0.071 ) & 0.965 & 0.300 & 0.308 ( 0.078 ) & 0.965 & 0.325 & 0.507 ( 0.050 ) & 0.966 & 0.209 & 0.299 ( 0.052 ) & 0.963 & 0.226\\
 & IPW & 2-probit & 0.488 ( 0.077 ) & 0.975 & 0.345 & 0.297 ( 0.092 ) & 0.963 & 0.390 & 0.480 ( 0.061 ) & 0.977 & 0.254 & 0.286 ( 0.072 ) & 0.951 & 0.283\\
 & IPW & 5-logit & 0.510 ( 0.090 ) & 0.947 & 0.370 & 0.308 ( 0.089 ) & 0.961 & 0.385 & 0.500 ( 0.068 ) & 0.933 & 0.261 & 0.296 ( 0.068 ) & 0.950 & 0.262\\
 & IPW & 5-probit & 0.492 ( 0.076 ) & 0.975 & 0.351 & 0.302 ( 0.094 ) & 0.964 & 0.407 & 0.483 ( 0.054 ) & 0.975 & 0.255 & 0.292 ( 0.070 ) & 0.954 & 0.294\\
 & IPW & 8-logit & 0.523 ( 0.070 ) & 0.958 & 0.292 & 0.310 ( 0.074 ) & 0.952 & 0.291 & 0.510 ( 0.048 ) & 0.951 & 0.197 & 0.300 ( 0.049 ) & 0.951 & 0.195\\
 & IPW & 8-probit & 0.492 ( 0.076 ) & 0.976 & 0.352 & 0.302 ( 0.094 ) & 0.965 & 0.407 & 0.483 ( 0.054 ) & 0.974 & 0.255 & 0.292 ( 0.070 ) & 0.955 & 0.294\\
\addlinespace
common $\tau_k=0.15$  & MRE &  & 0.575 ( 0.086 ) & 0.835 & 0.318 & 0.337 ( 0.089 ) & 0.899 & 0.321 & 0.571 ( 0.061 ) & 0.766 & 0.224 & 0.334 ( 0.06 ) & 0.892 & 0.225\\
 & IPW & 2-logit & 0.518 ( 0.087 ) & 0.938 & 0.332 & 0.305 ( 0.096 ) & 0.930 & 0.358 & 0.511 ( 0.061 ) & 0.937 & 0.237 & 0.300 ( 0.066 ) & 0.942 & 0.254\\
 & IPW & 2-probit & 0.489 ( 0.097 ) & 0.943 & 0.383 & 0.292 ( 0.113 ) & 0.931 & 0.422 & 0.478 ( 0.068 ) & 0.957 & 0.289 & 0.284 ( 0.087 ) & 0.932 & 0.312\\
 & IPW & 5-logit & 0.515 ( 0.111 ) & 0.930 & 0.408 & 0.304 ( 0.112 ) & 0.921 & 0.418 & 0.496 ( 0.098 ) & 0.904 & 0.301 & 0.294 ( 0.092 ) & 0.916 & 0.298\\
 & IPW & 5-probit & 0.492 ( 0.105 ) & 0.945 & 0.391 & 0.297 ( 0.117 ) & 0.944 & 0.442 & 0.482 ( 0.066 ) & 0.971 & 0.293 & 0.290 ( 0.091 ) & 0.935 & 0.326\\
 & IPW & 8-logit & 0.534 ( 0.084 ) & 0.934 & 0.323 & 0.309 ( 0.090 ) & 0.929 & 0.321 & 0.516 ( 0.058 ) & 0.924 & 0.222 & 0.301 ( 0.059 ) & 0.936 & 0.218\\
 & IPW & 8-probit & 0.492 ( 0.105 ) & 0.945 & 0.392 & 0.297 ( 0.117 ) & 0.941 & 0.442 & 0.482 ( 0.066 ) & 0.966 & 0.293 & 0.290 ( 0.091 ) & 0.935 & 0.325\\
\addlinespace
common $\tau_k=0.30$ & MRE &  & 0.622 ( 0.122 ) & 0.73 & 0.413 & 0.357 ( 0.12 ) & 0.877 & 0.423 & 0.621 ( 0.081 ) & 0.638 & 0.299 & 0.357 ( 0.085 ) & 0.853 & 0.3\\
& IPW & 2-logit & 0.531 ( 0.123 ) & 0.896 & 0.417 & 0.308 ( 0.129 ) & 0.913 & 0.450 & 0.523 ( 0.084 ) & 0.927 & 0.308 & 0.300 ( 0.095 ) & 0.907 & 0.323\\
 & IPW & 2-probit & 0.493 ( 0.135 ) & 0.924 & 0.480 & 0.290 ( 0.156 ) & 0.911 & 0.516 & 0.478 ( 0.095 ) & 0.955 & 0.369 & 0.277 ( 0.128 ) & 0.862 & 0.376\\
 & IPW & 5-logit & 0.527 ( 0.167 ) & 0.870 & 0.535 & 0.305 ( 0.155 ) & 0.896 & 0.537 & 0.496 ( 0.141 ) & 0.870 & 0.400 & 0.289 ( 0.133 ) & 0.864 & 0.383\\
 & IPW & 5-probit & 0.500 ( 0.138 ) & 0.925 & 0.496 & 0.296 ( 0.162 ) & 0.905 & 0.541 & 0.481 ( 0.093 ) & 0.961 & 0.384 & 0.283 ( 0.137 ) & 0.851 & 0.394\\
 & IPW & 8-logit & 0.562 ( 0.117 ) & 0.877 & 0.405 & 0.312 ( 0.120 ) & 0.896 & 0.399 & 0.535 ( 0.080 ) & 0.896 & 0.283 & 0.303 ( 0.083 ) & 0.901 & 0.274\\
& IPW & 8-probit & 0.500 ( 0.138 ) & 0.926 & 0.495 & 0.296 ( 0.162 ) & 0.903 & 0.540 & 0.481 ( 0.093 ) & 0.966 & 0.384 & 0.283 ( 0.137 ) & 0.847 & 0.394\\
\bottomrule
\end{tabular}}
\begin{tablenotes}
AVE, mean value of estimates; SD, standard error of estimates; CP, 95\%confidence interval coverage probability; LOCI, length of confidence interval; MRE, multivariate random-effects model without considering the publication bias; IPW, the proposed method using Inverse probability weighting technique; 2-logit, IPW estimator considering the 2-parameter logistic selection model; 2-probit, IPW estimator considering the 2-parameter probit selection model; 5-logit, IPW estimator considering the 5-parameter logistic selection model; 5-probit, IPW estimator considering the 5-parameter probit selection model
\end{tablenotes}
\end{table}
\end{center}

\begin{table}[h]

\caption{Simulation results for the heterogeneity estimation under the 5-parameter logistic selection function (Moderate-size)}
\centering
\resizebox{\linewidth}{!}{
\begin{tabular}[t]{llcccccccccccc}
\toprule
\multicolumn{1}{c}{} & \multicolumn{1}{c}{} & \multicolumn{3}{c}{$\tau_1$} & \multicolumn{3}{c}{$\tau_2$} & \multicolumn{3}{c}{$\tau_3$} & \multicolumn{3}{c}{$\tau_4$} \\
\cmidrule(l{3pt}r{3pt}){3-5} \cmidrule(l{3pt}r{3pt}){6-8} \cmidrule(l{3pt}r{3pt}){9-11} \cmidrule(l{3pt}r{3pt}){12-14}
Method & Selection & AVE(SD) & CP & NOZ & AVE(SD) & CP & NOZ & AVE(SD) & CP & NOZ & AVE(SD) & CP & NOZ\\
\midrule
MRE &  & 0.05 ( 0.095 ) & 0.997 & 491 & 0.076 ( 0.123 ) & 0.998 & 418 & 0.174 ( 0.198 ) & 0.987 & 251 & 0.281 ( 0.161 ) & 0.945 & 108\\
IPW & 2-logit & 0.051 ( 0.105 ) & 0.977 & 438 & 0.082 ( 0.130 ) & 1.000 & 368 & 0.164 ( 0.185 ) & 0.996 & 239 & 0.290 ( 0.151 ) & 0.915 & 40\\
IPW & 2-probit & 0.064 ( 0.149 ) & 0.982 & 434 & 0.095 ( 0.152 ) & 0.997 & 345 & 0.167 ( 0.187 ) & 0.996 & 242 & 0.300 ( 0.154 ) & 0.942 & 47\\
IPW & 5-logit & 0.065 ( 0.151 ) & 0.967 & 438 & 0.083 ( 0.133 ) & 1.000 & 385 & 0.159 ( 0.182 ) & 0.996 & 250 & 0.295 ( 0.154 ) & 0.929 & 49\\
IPW & 5-probit & 0.063 ( 0.148 ) & 0.984 & 427 & 0.094 ( 0.152 ) & 0.995 & 342 & 0.160 ( 0.187 ) & 0.995 & 253 & 0.303 ( 0.154 ) & 0.947 & 38\\
IPW & 8-logit & 0.048 ( 0.095 ) & 0.964 & 444 & 0.077 ( 0.121 ) & 0.998 & 374 & 0.149 ( 0.175 ) & 0.991 & 273 & 0.286 ( 0.150 ) & 0.908 & 47\\
IPW & 8-probit & 0.063 ( 0.148 ) & 0.989 & 427 & 0.094 ( 0.152 ) & 0.996 & 342 & 0.160 ( 0.187 ) & 0.996 & 253 & 0.303 ( 0.154 ) & 0.943 & 38\\
&True&0.05&&&0.15&&&0.20&&&0.30&&\\
\addlinespace
MRE &  & 0.031 ( 0.08 ) & 0.998 & 742 & 0.028 ( 0.082 ) & 0.999 & 773 & 0.039 ( 0.098 ) & 0.999 & 728 & 0.039 ( 0.082 ) & 0.983 & 698\\
IPW & 2-logit & 0.046 ( 0.100 ) & 0.982 & 483 & 0.045 ( 0.101 ) & 0.951 & 500 & 0.052 ( 0.107 ) & 0.910 & 461 & 0.051 ( 0.090 ) & 0.986 & 401\\
IPW & 2-probit & 0.060 ( 0.124 ) & 0.989 & 450 & 0.058 ( 0.142 ) & 0.978 & 498 & 0.056 ( 0.121 ) & 0.939 & 486 & 0.058 ( 0.099 ) & 0.986 & 381\\
IPW & 5-logit & 0.051 ( 0.107 ) & 0.983 & 449 & 0.047 ( 0.112 ) & 0.973 & 517 & 0.048 ( 0.105 ) & 0.942 & 505 & 0.056 ( 0.094 ) & 0.981 & 371\\
IPW & 5-probit & 0.059 ( 0.119 ) & 0.993 & 435 & 0.054 ( 0.130 ) & 0.989 & 488 & 0.053 ( 0.117 ) & 0.954 & 499 & 0.060 ( 0.103 ) & 0.990 & 385\\
IPW & 8-logit & 0.043 ( 0.092 ) & 0.959 & 493 & 0.040 ( 0.092 ) & 0.921 & 507 & 0.046 ( 0.098 ) & 0.878 & 507 & 0.049 ( 0.087 ) & 0.957 & 403\\
IPW & 8-probit & 0.059 ( 0.119 ) & 0.994 & 435 & 0.054 ( 0.130 ) & 0.986 & 488 & 0.053 ( 0.117 ) & 0.951 & 499 & 0.060 ( 0.103 ) & 0.987 & 385\\
&True&0.05&&&0.05&&&0.05&&&0.05&&\\
\addlinespace
MRE &  & 0.079 ( 0.12 ) & 0.992 & 484 & 0.069 ( 0.119 ) & 0.997 & 506 & 0.113 ( 0.161 ) & 0.996 & 439 & 0.106 ( 0.123 ) & 0.979 & 399\\
IPW & 2-logit & 0.089 ( 0.130 ) & 0.999 & 344 & 0.081 ( 0.128 ) & 1.000 & 387 & 0.116 ( 0.153 ) & 0.995 & 302 & 0.117 ( 0.123 ) & 0.986 & 224\\
IPW & 2-probit & 0.103 ( 0.148 ) & 0.996 & 330 & 0.094 ( 0.151 ) & 0.997 & 389 & 0.121 ( 0.169 ) & 0.994 & 327 & 0.128 ( 0.131 ) & 0.986 & 199\\
IPW & 5-logit & 0.098 ( 0.138 ) & 0.997 & 347 & 0.084 ( 0.133 ) & 0.999 & 387 & 0.111 ( 0.151 ) & 0.991 & 314 & 0.126 ( 0.131 ) & 0.985 & 202\\
IPW & 5-probit & 0.104 ( 0.154 ) & 0.996 & 339 & 0.096 ( 0.153 ) & 0.996 & 379 & 0.115 ( 0.163 ) & 0.994 & 324 & 0.131 ( 0.130 ) & 0.990 & 186\\
IPW & 8-logit & 0.087 ( 0.123 ) & 1.000 & 342 & 0.076 ( 0.119 ) & 0.998 & 391 & 0.103 ( 0.144 ) & 0.990 & 326 & 0.116 ( 0.124 ) & 0.979 & 234\\
IPW & 8-probit & 0.104 ( 0.154 ) & 0.996 & 339 & 0.096 ( 0.153 ) & 0.996 & 379 & 0.115 ( 0.163 ) & 0.993 & 324 & 0.131 ( 0.130 ) & 0.986 & 186\\
&True&0.15&&&0.15&&&0.15&&&0.15&&\\
\addlinespace
MRE &  & 0.171 ( 0.163 ) & 0.884 & 213 & 0.165 ( 0.181 ) & 0.875 & 273 & 0.282 ( 0.249 ) & 0.942 & 164 & 0.284 ( 0.162 ) & 0.934 & 83\\
IPW & 2-logit & 0.186 ( 0.190 ) & 0.920 & 206 & 0.168 ( 0.181 ) & 0.909 & 250 & 0.264 ( 0.229 ) & 0.963 & 159 & 0.287 ( 0.155 ) & 0.908 & 50\\
IPW & 2-probit & 0.203 ( 0.218 ) & 0.954 & 205 & 0.183 ( 0.195 ) & 0.949 & 247 & 0.268 ( 0.238 ) & 0.976 & 161 & 0.299 ( 0.158 ) & 0.937 & 57\\
IPW & 5-logit & 0.195 ( 0.188 ) & 0.937 & 194 & 0.166 ( 0.180 ) & 0.934 & 268 & 0.250 ( 0.231 ) & 0.971 & 180 & 0.300 ( 0.162 ) & 0.930 & 51\\
IPW & 5-probit & 0.201 ( 0.216 ) & 0.961 & 209 & 0.179 ( 0.197 ) & 0.963 & 284 & 0.257 ( 0.236 ) & 0.981 & 168 & 0.304 ( 0.158 ) & 0.946 & 45\\
IPW & 8-logit & 0.181 ( 0.183 ) & 0.902 & 197 & 0.160 ( 0.172 ) & 0.871 & 254 & 0.237 ( 0.221 ) & 0.938 & 192 & 0.285 ( 0.154 ) & 0.899 & 61\\
IPW & 8-probit & 0.201 ( 0.216 ) & 0.963 & 209 & 0.179 ( 0.197 ) & 0.957 & 284 & 0.257 ( 0.236 ) & 0.972 & 168 & 0.304 ( 0.158 ) & 0.946 & 45\\
&True&0.30&&&0.30&&&0.30&&&0.30&&\\
\bottomrule
\end{tabular}}
\begin{tablenotes}
AVE, mean value of estimates; SD, standard error of estimates; CP, 95\%confidence interval coverage probability; NOZ, number of zero estimates; MRE, multivariate random-effects model without considering the publication bias; IPW, the proposed method using Inverse probability weighting technique; 2-logit, IPW estimator considering the 2-parameter logistic selection model; 2-probit, IPW estimator considering the 2-parameter probit selection model; 5-logit, IPW estimator considering the 5-parameter logistic selection model; 5-probit, IPW estimator considering the 5-parameter probit selection model
\end{tablenotes}
\end{table}

\begin{center}
\begin{table*}[h] 
\centering
\caption{The publication status of studies in the network meta-analysis of Antidepressant Drugs \label{pub.status}}
\begin{tabular}{lccc}
 \hline
 Study design  & Published & Unpublished & Proportion($\%$)\\
 \hline 
 Paroxetine VS Placebo & 35 & 12 &74.5\\
 Fluvoxamine VS Placebo & 25 & 12 &67.6\\
 Paroxetine VS Fluvoxamine & 8 & 2 &80.0\\
 Paroxetine VS Fluvoxamine VS Placebo & 1 & 2 &33.3\\
 All&69&28&71.1\\
\hline 
\end{tabular}
\end{table*}
\end{center}

\begin{center}
\begin{table*}[h] 
\centering
\caption{Summary of the statistical analysis for publication bias evaluation of the network meta-analysis of the Antidepressant Drugs\label{tab2}}
\begin{tabular}{llclcc}
\toprule
\textbf{Data} & \textbf{$\#$ of studies} & \textbf{Method}  & \textbf{Selection }  & \textbf{$\mu_1$}  & \textbf{$\mu_2$} \\
\midrule
Published $ \& $ unpublished  & $N^*$=81 &MRE & & 0.503 [0.413,0.593]  & 0.348 [0.215,0.481]     \\
Published & N = 69& MRE  &   & 0.594 [0.496,0.693]  & 0.411 [0.253,0.569]   \\
\hline
Published $\&$ Registry & N = 69, M = 28& IPW  & 2-logit  & 0.535 [0.416,0.657]  & 0.297 [0.114,0.488]   \\
Published $\&$ Registry & N = 69, M = 28& IPW  & 2-probit  & 0.527 [0.377,0.650]  & 0.213 [0.000,0.419]   \\
Published $\&$ Registry & N = 69, M = 28& IPW  & 5-logit  & 0.510 [0.324,0.648]  & 0.232 [-0.158,0.503]   \\
Published $\&$ Registry & N = 69, M = 28& IPW  & 5-probit  & 0.528 [0.370,0.664]  & 0.271 [-0.004,0.541]   \\
Published $\&$ Registry & N = 69, M = 28& IPW  & 8-logit  & 0.518 [0.401,0.633]  & 0.307 [0.127,0.491]   \\
Published $\&$ Registry & N = 69, M = 28& IPW  & 8-probit  & 0.488 [0.314,0.624]  & 0.273 [0.064,0.503]   \\
\bottomrule
\end{tabular}
\begin{tablenotes}
$\mu_1$, the estimates for the comparison of Paroxetine versus Placebo; $\mu_2$, the estimates for the comparison of Fluoxetine versus Placebo; MRE, multivariate random-effects model without considering the publication bias; IPW, the proposed method using Inverse probability weighting technique; 2-logit, IPW estimator considering the 2-parameter logistic selection model; 2-probit, IPW estimator considering the 2-parameter probit selection model; 5-logit, IPW estimator considering the 5-parameter logistic selection model; 5-probit, IPW estimator considering the 5-parameter probit selection model; 8-logit, IPW estimator considering the 8-parameter logistic selection model; 8-probit, IPW estimator considering the 8-parameter probit selection model
\end{tablenotes}
\end{table*}
\end{center}

\begin{center}
\begin{sidewaystable}[h] 
\centering
\caption{Summary of the statistical analysis for heterogeneity estimation in the network meta-analysis of the Antidepressant Drugs\label{tab2}}
\scalebox{0.85}[1]{
\begin{tabular*}{700pt}
{@{\extracolsep\fill}llclcccc@{\extracolsep\fill}}
\toprule
\textbf{Data} &\textbf{$\#$ of studies} & \textbf{Method}  & \textbf{Selection }  & \textbf{$\tau_1$}  & \textbf{$\tau_2$}& \textbf{$\tau_3$}& \textbf{$\tau_4$} \\
\midrule
Published $\&$ unpublished & $N^*$=81& MRE & & 0.086 [0.000,0.295]  & 0.334 [0.171,0.497] &0.000 [0.000,0.450]&0.297 [0.035,0.559]    \\
Published & N = 69 & MRE  &   & 0.000 [0.000,0.190]  & 0.382 [0.174,0.589]&0.000 [0.000,0.551]&0.000 [0.000,0.711]   \\
\hline
Published $\&$ Registry & N = 69, M = 28& IPW  & 2-logit  & 0.000 [0.000,0.202]  & 0.443 [0.215,0.687] &0.000 [0.000,0.394]& 0.000 [0.000,0.688] \\
Published $\&$ Registry & N = 69, M = 28& IPW  & 2-probit  & 0.000 [0.000,0.235]  & 0.483 [0.277,0.725] &0.000 [0.000,0.457]& 0.000 [0.000,0.691]  \\
Published $\&$ Registry & N = 69, M = 28& IPW  & 5-logit  & 0.000 [0.000,0.251]  & 0.432 [0.185,0.752] &0.000 [0.000,0.604]& 0.000 [0.000,0.747]   \\
Published $\&$ Registry & N = 69, M = 28& IPW  & 5-probit  & 0.000 [0.000,0.252]  & 0.466 [0.238,0.758] &0.000 [0.000,0.514]& 0.000 [0.000,0.729]   \\
Published $\&$ Registry & N = 69, M = 28& IPW  & 8-logit  & 0.000 [0.000,0.214]  & 0.421 [0.204,0.646] &0.000 [0.000,0.309]& 0.000 [0.000,0.691]  \\
Published $\&$ Registry & N = 69, M = 28& IPW  & 8-probit  & 0.000 [0.000,0.343]  & 0.467 [0.263,0.694] &0.000 [0.000,0.301]& 0.000 [0.000,0.697]   \\
\bottomrule
\end{tabular*}}
\begin{tablenotes}
$\mu_1$, the estimates for the comparison of Paroxetine versus Placebo; $\mu_2$, the estimates for the comparison of Fluoxetine versus Placebo; MRE, multivariate random-effects model without considering the publication bias; IPW, the proposed method using Inverse probability weighting technique; 2-logit, IPW estimator considering the 2-parameter logistic selection model; 2-probit, IPW estimator considering the 2-parameter probit selection model; 5-logit, IPW estimator considering the 5-parameter logistic selection model; 5-probit, IPW estimator considering the 5-parameter probit selection model; 8-logit, IPW estimator considering the 8-parameter logistic selection model; 8-probit, IPW estimator considering the 8-parameter probit selection model
\end{tablenotes}
\end{sidewaystable}
\end{center}

\begin{center}
\begin{table*}[h] 
\centering
\caption{Comparison of $P$-scores for ranking treatments in the network meta-analysis of the Antidepressant Drugs with and without considering the publication bias\label{tab2}}%
\begin{tabular*}{500pt}{@{\extracolsep\fill}llclcc@{\extracolsep\fill}}
\toprule

\textbf{Data} &\textbf{$\#$ of studies}& \textbf{Method}  & \textbf{Selection }  & \textbf{$P_{Paroxetine}$}  & \textbf{$P_{Fluoxetine}$} \\
\midrule
Published $\&$ unpublished  & $N^*$=81& MRE & & 0.991  & 0.509      \\
Published & N = 69& MRE  &   & 0.994  & 0.506   \\
\hline
Published $\&$ Registry &N = 69, M = 28& IPW  & 2-logit  & 0.995  & 0.505   \\
Published $\&$ Registry &N = 69, M = 28& IPW  & 2-probit  & 0.999   & 0.493   \\
Published $\&$ Registry &N = 69, M = 28& IPW  & 5-logit  & 0.967   & 0.484   \\
Published $\&$ Registry &N = 69, M = 28& IPW  & 5-probit  & 0.995   & 0.481   \\
Published $\&$ Registry &N = 69, M = 28& IPW  & 8-logit  & 0.993   & 0.507   \\
Published $\&$ Registry &N = 69, M = 28& IPW  & 8-probit  & 0.990   & 0.508   \\
\bottomrule
\end{tabular*}
\begin{tablenotes}
$P_{Paroxetine}$, P-scores for the treatment of paroxetine; $P_{Fluoxetine}$, P-scores for the treatment of Fluoxetine; MRE, multivariate random-effects model without considering the publication bias; IPW, the proposed method using Inverse probability weighting technique; 2-logit, IPW estimator considering the 2-parameter logistic selection model; 2-probit, IPW estimator considering the 2-parameter probit selection model; 5-logit, IPW estimator considering the 5-parameter logistic selection model; 5-probit, IPW estimator considering the 5-parameter probit selection model; 8-logit, IPW estimator considering the 8-parameter logistic selection model; 8-probit, IPW estimator considering the 8-parameter probit selection model
\end{tablenotes}
\end{table*}
\end{center}

\end{document}